\documentclass[epjc3]{svjour3} 

\usepackage[margin=25mm]{geometry}
\usepackage{amsmath,amssymb}

\newcommand{\Sq}[1]{S_q^{(#1)}}
\newcommand{\Xq}[1]{X_q^{(#1)}}
\newcommand{\Uq}[1]{U_q^{(#1)}}
\newcommand{\Vol}[1]{V^{(#1)}}

\newcommand{\TS}{T^{(S)}}%
\newcommand{\PS}{P^{(S)}}%

\newcommand{\Tph}{T_{\mathrm{ph}}}
\newcommand{\Pph}{P_{\mathrm{ph}}}
\newcommand{\TphS}{T_{\mathrm{ph}}^{(S)}}
\newcommand{\PphS}{P_{\mathrm{ph}}^{(S)}}
\newcommand{\TphE}{T_{\mathrm{ph}}^{(E)}}
\newcommand{\PphE}{P_{\mathrm{ph}}^{(E)}}
\newcommand{\TphB}{T_{\mathrm{ph}}^{(B)}}
\newcommand{\PphB}{P_{\mathrm{ph}}^{(B)}}
\newcommand{\Heat}[1]{Q^{(#1)}}%
\newcommand{\tTphS}{\tilde{T}_{\mathrm{ph}}^{(S)}}
\newcommand{\tPphS}{\tilde{P}_{\mathrm{ph}}^{(S)}}

\newcommand{\kappaTS}{\kappa_{T}^{(S)}}
\newcommand{\kappaXS}{\kappa_{X}^{(S)}}
\newcommand{\etaphS}{\eta_{\mathrm{ph}}^{(S)}}

\newcommand{\FreeEne}[1]{F_q^{(#1)}}%

\newcommand{\pxderiv}[2]{\Bigg(\frac{\partial #1}{\partial #2}\Bigg)}

\def\<#1>{\langle #1 \rangle}

\usepackage{color}

\begin{document}
\title{Thermodynamic relations and fluctuations in the Tsallis statistics}
\author{Masamichi Ishihara\thanksref{e1,addr1}}
\thankstext{e1}{email: m\_isihar@koriyama-kgc.ac.jp}
\institute{Department of Food and Nutrition, Koriyama Women's University, Koriyama, Fukushima, 963-8503, Japan \label{addr1}}

\abstractdc{
  The thermodynamic relations in the Tsallis statistics were studied with physical quantities. 
  An additive entropic variable related to the Tsallis entropy was introduced by assuming the form of the first law of the thermodynamics.
  The fluctuations in the Tsallis statistics were derived with physical quantities with the help of the introduced entropic variable.
  It was shown that the mean squares of the fluctuations of the physical quantities in the Tsallis statistics
  are the same as those in the conventional statistics. 
  The mean square of the fluctuation of the Tsallis entropy and the mean square of the fluctuation of the Tsallis temperature were also derived.
  The mean square of the relative fluctuation of the Tsallis entropy and
  the mean square of the relative fluctuation of the Tsallis temperature are represented with heat capacities.
  It was shown that these fluctuations of the Tsallis quantities have the $q$-dependent terms in the Tsallis statistics of the entropic parameter $q$.
}

\maketitle

\section{Introduction}

The statistics which show power-like distributions have been interested in many branches of science.
One of them is the Tsallis statistics which is an possible extension of the Boltzmann-Gibbs statistics, 
and the statistics has been applied in various fields \cite{TsallisBook,Tsallis:Entropy:2019}.
The entropy called Tsallis entropy and the escort average are employed in this statistics,
and the probability distribution is obtained in the maximum entropy principle.
The relations between thermodynamic quantities, such as internal energy and entropy, have been discussed. 
The statistics may describe the phenomena which show power-like distributions.

The physical temperature and the physical pressure were introduced with the Tsallis entropy
\cite{Kalyana:2000,Abe-PLA:2001,S.Abe:physicaA:2001,Aragao:2003,Ruthotto:2003,Toral:2003,Suyari:2006}.
In the Boltzmann-Gibbs statistics,
the inverse temperature is given by the partial derivative of the entropy with respect to the internal energy.
In the Tsallis statistics, the physical temperature was introduced in the similar way,
though the inverse temperature-like parameter appears as a Lagrange multiplier in the maximum entropy principle.
The physical temperature seems to be an appropriate variable to describe the system \cite{Ishihara:phi4,Ishihara:free-field,Ishihara:2023}.

An entropic variable as a function of the Tsallis entropy was introduced by
considering the Legendre transform structure in the Tsallis statistics \cite{S.Abe:physicaA:2001}. 
It is considered that the Legendre transform structure is an essential ingredient \cite{plastino1997}.
In contrast, it is rarely noted that 
the Legendre transform structure may be unnecessary in the unconventional statistics \cite{C-Yepes}.
It was also shown that the Legendre transform structure is robust against the choice of entropy and the definition of mean value
\cite{plastino1997,yamano2000}. 
The variables can not be determined uniquely by the Legendre transform structure,
though the structure determines the conjugate variable for a given variable \cite{S.Abe:physicaA:2001}.
Therefore, it may be better to introduce the entropic variable without using the Legendre transform structure explicitly
when the variables can be determined by another consideration, though the structure is desirable.
The entropic variable is useful in the description of the thermodynamics,
because a temperature-like parameter is defined by using the entropy
and because the Legendre transform structure is supported by the relation between the entropic variable and the temperature-like parameter. 

It is considered that Tsallis-type distributions are related to fluctuations.
The Tsallis-type distributions are often used to describe the phenomena,
such as the momentum distributions \cite{Cleymans2012,Marques2015,Azmi2015,Thakur2016,Khuntia2017,Si2017,Bhattacharyya2018,Parvan2020}
and fluctuations \cite{Osada:Isihara:2018} at high energies.
The distribution was obtained by assuming that the heat capacity of the environment is exactly constant \cite{Watanabe2004}.
The entropic parameter $q$ of the Tsallis statistics is related to the heat capacity
which is connected with the fluctuation of the inverse temperature \cite{Wilk2009}.
The relation between the fluctuation and the entropic parameter was discussed in the study of the time dependence of the entropic parameter \cite{Wilk2021}.
This distribution was obtained for the system with fluctuation \cite{Wilk2000,Saha2021}.
The distribution was derived in the studies of
critical end point \cite{Ayala2020}, quantum entanglement \cite{Ourabah2017,Castano2021}, entropy exchange \cite{Castano2022}, and so on.
The parameter $q$ is also related to the number of the configurations of intensive quantity \cite{Castano2021}.
The property of the parameter $q$ should be important in the Tsallis statistics \cite{Parvan2006}.

The fluctuations of the thermodynamic quantities are significant in the Tsallis statistics.
The fluctuation of the energy in the canonical ensemble was obtained \cite{Liyan:2008}
by solving the differential equation for $0<q<1$ in the optimal Lagrange multiplier formalism \cite{Martinez2000}.
The fluctuations were also calculated by maximizing the entropy that is constructed from probabilities
with the deviation parameter from the equilibrium value \cite{Vives:2002}.
The calculations of the physical quantities and the calculations of the Tsallis quantities are required to clarify the effects of the statistics.

In this paper, we consider thermodynamic relations,
and attempt to find the expressions of the fluctuations in the Tsallis statistics.  
In section \ref{sec:thermodyn},
we consider the thermodynamic relations with physical quantities,
such as physical temperature and physical pressure.
In section \ref{sec:fluctuation},
the fluctuations are discussed in the Tsallis statistics with the introduced entropic variable.
The fluctuations of the physical quantities and the fluctuations of the Tsallis quantities are obtained. 
The last section is assigned for conclusions.   

\section{Thermodynamic relations with physical quantities}
\label{sec:thermodyn} 
We treat a system and an environment.
The system and the environment are labeled with the superscripts $(S)$ and $(E)$, respectively.
The total system constructed from the system and the environment is labeled by the superscript $(S+E)$.

We attempt to find the relations among
the internal energy $U_q$, the physical temperature $\Tph$, the entropic variable $X_q$,
the physical pressure $\Pph$, and the volume $V$.
The entropic variable $X_q$ was already introduced in the reference~\cite{S.Abe:physicaA:2001}.
This variable $X_q$ is given below in this paper. 
The following discussion is based on the discussion given in the references \cite{Abe-PLA:2001} and \cite{S.Abe:physicaA:2001}.

The Tsallis entropy $S_q(U_q, V)$ with the entropic parameter $q$ satisfies the following relation:
\begin{align}
  \Sq{S+E} = \Sq{S} + \Sq{E} + (1-q) \Sq{S} \Sq{E} .
  \label{eqn:entropy}
\end{align}
The additivity of the internal energy is assumed:
\begin{align}
  \Uq{S+E} = \Uq{S} + \Uq{E} .
  \label{eqn:energy}
\end{align}
The total volume $\Vol{S+E}$ is the sum of the volumes, $\Vol{S}$ and $\Vol{E}$:
\begin{align}
  \Vol{S+E} = \Vol{S} + \Vol{E} .
  \label{eqn:volume}
\end{align}

The maximum entropy principle requires $\delta \Sq{S+E} = 0$,
and the total internal energy and the total volume satisfy $\delta \Uq{S+E} = 0$ and $\delta \Vol{S+E}=0$.  
With these requirements, we define the physical temperature $\Tph$ and the physical pressure $\Pph$:
\begin{subequations}
\begin{align}
  &\frac{1}{\TphS}
  = \frac{1}{1+(1-q)\Sq{S}} \left( \frac{\partial \Sq{S}}{\partial \Uq{S}} \right)_{\Vol{S}},
  \label{def:Tph}\\
  &\frac{1}{\TphE}
  = \frac{1}{1+(1-q)\Sq{E}} \left( \frac{\partial \Sq{E}}{\partial \Uq{E}} \right)_{\Vol{E}},
  \\ 
  &\frac{\PphS}{\TphS} 
  = \frac{1}{1+(1-q)\Sq{S}} \left( \frac{\partial \Sq{S}}{\partial \Vol{S}} \right)_{\Uq{S}}, 
  \label{def:Pph}\\
  &\frac{\PphE}{\TphE} 
  = \frac{1}{1+(1-q)\Sq{E}} \left( \frac{\partial \Sq{E}}{\partial \Vol{E}} \right)_{\Uq{E}}.
\end{align}
\end{subequations}
We have the relations $\TphS = \TphE$ and $\PphS = \PphE$
from Eqs.\eqref{eqn:entropy}, \eqref{eqn:energy}, and \eqref{eqn:volume} with these definitions. 
These equations $\TphS = \TphE$ and $\PphS = \PphE$ indicate that
the physical temperature and the physical pressure characterize the equilibrium. 
We use the names, physical temperature and physical pressure, in this paper,
though names, equilibrium temperature and equilibrium pressure, might be adequate for the above introduced temperature and pressure.

The differential of the Tsallis entropy is 
\begin{align}
  d\Sq{S} = \left( \frac{\partial \Sq{S}}{\partial \Uq{S}} \right)_{\Vol{S}} d\Uq{S}
    + \left( \frac{\partial \Sq{S}}{\partial \Vol{S}} \right)_{\Uq{S}} d\Vol{S}
    .
\end{align}
We have the following relation by using Eqs.~\eqref{def:Tph} and \eqref{def:Pph}:
\begin{align}
  d\Uq{S} = \Bigg( \frac{\TphS}{1+(1-q)\Sq{S}} \Bigg) d\Sq{S} - \Pph d\Vol{S}.
\label{Tsallis:firstlaw}
\end{align}
This is the first law of the thermodynamics in the Tsallis statistics.
We introduce an entropic variable $\Xq{S}$ by requiring that Eq.~\eqref{Tsallis:firstlaw} has the following form: 
\begin{align}
  d\Uq{S} = \TphS d\Xq{S} - \PphS d\Vol{S}.
\label{eqn:1st-law:X}
\end{align}
This requirement is satisfied by defining $\Xq{S}$ as
\begin{align}
\Xq{S} = \frac{1}{1-q} \ln(1+(1-q)\Sq{S}). 
\label{eqn:relation:XS}
\end{align}
The entropic variables $\Xq{E}$ and $\Xq{S+E}$ are defined in the same way.
From Eq.~\eqref{eqn:1st-law:X}, it is natural to define the heat transfer $\Heat{S}$ as follows:
\begin{align}
  \Heat{S} =  \TphS d\Xq{S}.
\end{align}

The alternative definition of heat transfer is given as $T^{(S)} d\Sq{S}$
by using the temperature $T^{(S)}$ which is the inverse of the Lagrange multiplier \cite{C.Tsallis1998}.
The temperature $\TS$ is called the Tsallis temperature in this paper.
It may be worth to mention that the relation, $\TphS d\Xq{S} = T^{(S)} d\Sq{S}$, is easily shown \cite{Ishihara:EPJB:95}. 
Similar relation between the heat transfer in the incomplete non-extensive statistics
and that in the R\'enyi statistics was shown in the reference \cite{Parvan2004}.

As pointed by some researchers \cite{S.Abe:physicaA:2001,Wang:prepri:2003}, 
the introduced entropic variables, $\Xq{S}$ and $\Xq{E}$, are additive:
\begin{align}
\Xq{S+E} = \Xq{S} + \Xq{E}.
\end{align}
This property is easily shown from the pseudo-additivity of the Tsallis entropy.  
The pseudo-additivity of the Tsallis entropy $S_q$ is mapped to the additivity of the entropic variable $X_q$.

Equation~\eqref{eqn:1st-law:X} indicates that the variables $\TphS$ and  $\Xq{S}$ are a Legendre pair.
Therefore,
the free energy $\FreeEne{S}$ in terms of $\TphS$ is naturally introduced by using the Legendre transformation of $\Uq{S}$
\cite{Abe-PLA:2001,Ishihara:EPJB:95}: 
\begin{align}
\FreeEne{S} = \Uq{S} - \TphS \Xq{S}. 
\end{align}
There is another definition of the free energy $\tilde{F}^{(S)}$ \cite{C.Tsallis1998,Ishihara:EPJB:95}
which is given by $\tilde{F}^{(S)} = \Uq{S} - T^{(S)} \Sq{S}$.

The entropic variable $X_q$ is valid for the physical temperature $\Tph$
because of the relation among $\Tph$, $X_q$, and $U_q$: $1/\Tph = (\partial X_q/\partial U_q)_V$. 
The first law and the heat transfer are described with $\Tph$ and $X_q$, 
and the free energy is defined by using Legendre transformation with $\Tph$ and $X_q$, as shown above.

\section{Fluctuations in the Tsallis statistics}
\label{sec:fluctuation}
\subsection{The entropies and the number of states}
The Tsallis entropies \cite{TsallisBook,S.Abe:PRE:2002,Moyano:EurLett:73} are given by 
\begin{subequations}
\begin{align}
  &\Sq{S} = \ln_q W^{(S)} , \label{def:Sq:S:micro}\\
  &\Sq{E} = \ln_q W^{(E)} , \label{def:Sq:E:micro}\\
  &\Sq{S+E} = \ln_q W^{(S+E)} , \label{def:Sq:SE:micro} 
\end{align}
\end{subequations}
where $W$ represents the number of states and $\ln_q x$ is the $q$-logarithm function.
As for the number of states, 
we assume that the system $S$ and the environment $E$ are independent:
\begin{align}
W^{(S+E)} = W^{(S)} W^{(E)} .
\end{align}
In such a case, the Tsallis entropy has the pseudo-additivity which is shown from
Eqs.~\eqref{def:Sq:S:micro}, \eqref{def:Sq:E:micro}, and \eqref{def:Sq:SE:micro}:
\begin{align}
  \Sq{S+E} &= \Sq{S} + \Sq{E} + (1-q) \Sq{S} \Sq{E}.
\end{align}
By substituting Eq.~\eqref{def:Sq:SE:micro} into the definition of $\Xq{S+E}$, $\Xq{S+E} = (1-q)^{-1} \ln(1+(1-q) \Sq{S+E})$, 
the entropic variable $\Xq{S+E}$ has the following relation:
\begin{align}
  W^{(S+E)} = \exp (\Xq{S+E}) . 
\label{W-Xq-Rel}
\end{align}
We estimate fluctuations by using Eq.~\eqref{W-Xq-Rel}.

The entropic variable $X_q$ is given by $X_q=\ln W$ as shown above, where we omit the superscript.
As is well-known, the Boltzmann-Gibbs entropy $S_{\mathrm{BG}}$ is given by $S_{\mathrm{BG}} = \ln W$:
we have $X_q = S_{\mathrm{BG}}$.
Therefore, the quantity derived from $X_q$ coincides with the quantity derived from $S_{\mathrm{BG}}$.
For example, $1/T_{\mathrm{ph}} = (\partial X_q/\partial U_q)_V = (\partial S_{\mathrm{BG}}/\partial U_q)_V = 1/T_{\mathrm{BG}}$, 
where $T_{\mathrm{BG}}$ is the temperature in the Boltzmann-Gibbs statistics.

\subsection{Fluctuations of the physical quantities}

Equation~\eqref{W-Xq-Rel} is the well-known form in the Boltzmann-Gibbs statistics.
We note calculations to clarify the procedure, though the following procedure is standard in the conventional statistics.
In the following calculations, we deal with the deviation $\Delta f$ of a function $f(x,y)$. 
The deviation $\Delta f$ is defined as $\Delta f = f(x+\Delta x, y+\Delta y) - f(x,y)$.
We introduce the quantity $\Delta \Xq{S+E}$ which is the deviation from the equilibrium value $\Xq{S+E}$ of the isolated system $S+E$,
and introduce the probability $P_r(\Delta \Xq{S+E})$ which is the probability of the occurrence of $\Delta \Xq{S+E}$.

The probability $P_r(\Delta \Xq{S+E})$ is given by
\begin{align}
  P_r(\Delta \Xq{S+E})
  = \frac{\exp (\Xq{S+E}+\Delta \Xq{S+E})}{\displaystyle\sum_{\Delta \Xq{S+E}} \exp (\Xq{S+E}+\Delta \Xq{S+E})}
  = \frac{\exp (\Delta \Xq{S+E})}{\displaystyle\sum_{\Delta \Xq{S+E}} \exp (\Delta \Xq{S+E})} . 
\end{align}
Therefore, we focus on $\Delta \Xq{S+E}$.

The quantity $\Delta \Xq{S+E}$ is given by
\begin{align}
  \Delta \Xq{S+E} &= \Delta \Xq{S} + \Delta \Xq{E} .
\end{align}
The energy $\Uq{S}$ and the volume $\Vol{S}$ fluctuate,
and the entropy $\Xq{S}$ as a function of $\Uq{S}$ and $\Vol{S}$ fluctuate.
The quantity $\Delta \Xq{S}$ is expanded
with Eqs.~\eqref{def:Tph}, \eqref{def:Pph}, and \eqref{eqn:relation:XS} as follows:
\begin{align}
  \Delta \Xq{S}(\Uq{S},\Vol{S}) 
  & = \Bigg( \frac{1}{\TphS}\Bigg) (\Delta \Uq{S})
  + \Bigg(\frac{\PphS}{\TphS} \Bigg) (\Delta \Vol{S})
  \nonumber \\ & \quad 
  + \frac{1}{2} \Bigg(\Delta \Bigg( \frac{1}{\TphS}\Bigg)\Bigg) (\Delta \Uq{S})
  + \frac{1}{2} \Bigg(\Delta\Bigg(\frac{\PphS}{\TphS}\Bigg)\Bigg) (\Delta \Vol{S})
  \nonumber \\ & \quad 
  + O(\Delta^3)
  .
\end{align}
The last term, $ O(\Delta^3)$, represents $(\Delta \Uq{S})^i (\Delta \Vol{S})^j$ terms $(i+j \ge 3)$.
In the same way, we obtain $\Delta \Xq{E}$:
\begin{align}
  \Delta \Xq{E}(\Uq{E},\Vol{E})
  &= \Bigg( \frac{1}{\TphE}\Bigg) (\Delta \Uq{E})
  + \Bigg(\frac{\PphE}{\TphE} \Bigg) (\Delta \Vol{E})
  \nonumber \\ & \quad 
  + \frac{1}{2} \Bigg(\Delta \Bigg( \frac{1}{\TphE}\Bigg)\Bigg) (\Delta \Uq{E})
  + \frac{1}{2} \Bigg(\Delta\Bigg(\frac{\PphE}{\TphE}\Bigg)\Bigg) (\Delta \Vol{E})
  \nonumber \\ & \quad   
  + O(\Delta^3)
  .
\label{XqE:expansion}
\end{align}

Hereafter, we treat the case that the environment is the bath with $\Delta \TphB = \Delta \PphB = 0$.
We attach the superscript $(B)$ for the bath instead of $(E)$. 
For the bath, from Eq.~\eqref{XqE:expansion}, the quantity $\Delta \Xq{B}$ is given by
\begin{align}
  \Delta \Xq{B}(\Uq{B},\Vol{B}) =  \Bigg( \frac{1}{\TphB}\Bigg) (\Delta \Uq{B})+ \Bigg(\frac{\PphB}{\TphB} \Bigg) (\Delta \Vol{B})
  + O(\Delta^3) .
\end{align}
The deviation $\Delta \Xq{S+B}$ with $\Delta \Uq{S+B}= \Delta \Vol{S+B} = 0$ is given by 
\begin{align}
  \Delta \Xq{S+B} &= \Delta \Xq{S}(\Uq{S},\Vol{S}) + \Delta \Xq{B}(\Uq{B},\Vol{B})
  \nonumber \\
  & = \frac{1}{2} \Bigg(\Delta \Bigg( \frac{1}{\TphS}\Bigg)\Bigg) (\Delta \Uq{S})
  + \frac{1}{2} \Bigg(\Delta\Bigg(\frac{\PphS}{\TphS}\Bigg)\Bigg) (\Delta \Vol{S})
  + O(\Delta^3)
  .
\label{XqSB:expansion:1}
\end{align}

We expand $\Delta \Uq{S}(\Xq{S}, \Vol{S})$ in order to represent the right-hand side of Eq.~\eqref{XqSB:expansion:1}
with the variables, $\Xq{S}$ and $\Vol{S}$:
\begin{align}
\Delta \Uq{S}(\Xq{S}, \Vol{S}) = \tTphS \Delta \Xq{S} - \tPphS \Delta \Vol{S} + O(\Delta^2),  
\label{Uqs:expansion}
\end{align}
where $\tTphS$ and $\tPphS$ are defined by
\begin{subequations}
\begin{align}
  &\tTphS = \left.\pxderiv{\Uq{S}}{\Xq{S}}\right._{\Vol{S}} ,\\
  &\tPphS = -\left.\pxderiv{\Uq{S}}{\Vol{S}}\right._{\Xq{S}} .
\end{align}
\end{subequations}
Substituting Eq.~\eqref{Uqs:expansion} into Eq.~\eqref{XqSB:expansion:1} with $\TphS = \tTphS$ and $\PphS=\tPphS$,
we have
\begin{align}
\Delta{\Xq{S+B}} = - \frac{1}{2\TphS} \left[ (\Delta \TphS) (\Delta \Xq{S}) - (\Delta \PphS) (\Delta \Vol{S}) \right] + O(\Delta^3) .
\end{align}
As a result, the probability $P_r(\Delta \Xq{S+B})$ is approximately given by 
\begin{subequations}
\begin{align}
  & P_r(\Delta{\Xq{S+B}}) = N^{-1} \exp\Bigg(- \frac{1}{2\TphS} \left[ (\Delta \TphS) (\Delta \Xq{S}) - (\Delta \PphS) (\Delta \Vol{S}) \right]\Bigg), 
\end{align}
where $N$ is the normalization constant.
We can choose convenient variables for calculations. 
When the physical temperature and the volume are adopted as variables, 
the constant $N$ is given by
\begin{align}
& \qquad N = \int_{D} d(\Delta \TphS) d(\Delta \Vol{S}) \ P_r(\Delta{\Xq{S+B}}) ,  
\end{align}
\end{subequations}
where the notation $D$ represents the appropriate region of the integral.
This region comes from the restrictions of the parameters.
For example, the volume of the system is not less than zero.

It is possible to calculate the mean squares of the fluctuations with the probability.
For example, the mean square of the fluctuation of the physical temperature is given by 
\begin{align}
  & \<(\Delta \TphS)^2> = N^{-1} \int_{D} d(\Delta \TphS) d(\Delta \Vol{S}) \ P_r(\Delta{\Xq{S+B}}) \ (\Delta \TphS)^2 .
\end{align}
As we obtain the mean squares of the fluctuations in the conventional statistics, we have
\begin{subequations}
\begin{align}
  \<(\Delta \TphS)^2> &\sim \frac{(\TphS)^2}{C_{qV}^{(S)}},\label{eq:fluc:Tph}\\ 
  \<(\Delta \Vol{S})^2> &\sim \TphS \kappaTS \Vol{S},\\ 
  \<(\Delta \Xq{S})^2> &\sim C_{qP}^{(S)} , \label{eq:fluc:X}\\ 
  \<(\Delta \PphS)^2> &\sim \frac{\TphS}{\Vol{S}\kappaXS}, 
\end{align}
\label{fluctuations}
\end{subequations}
where $C_{qV}^{(S)}$ is the heat capacity at constant volume, 
$C_{qP}^{(S)}$ is the heat capacity at constant (physical) pressure, 
$\kappaTS$ is the isothermal compressibility, 
$\kappaXS$ is the adiabatic compressibility, 
and $\Vol{S}$ is the volume of the system.
The heat capacities, $C_{qV}$ and $C_{qP}$, are given by 
\begin{subequations}
\begin{align}
  C_{qV} &= \Tph  \Bigg( \frac{\partial X_{q}}{\partial \Tph} \Bigg)_{V}
  = \Bigg( \frac{\partial U_q}{\partial \Tph} \Bigg)_{V} ,
  \label{eqn:CqV} \\
  C_{qP} &= \Tph  \Bigg( \frac{\partial X_{q}}{\partial \Tph} \Bigg)_{\Pph} . 
  \label{eqn:CqP}
\end{align}
The compressibilities, $\kappa_T$ and $\kappa_X$, are given by
\begin{align}
  \kappa_T &= -\frac{1}{V} \Bigg( \frac{\partial V}{\partial \Pph} \Bigg)_{\Tph} ,\\
  \kappa_X &= -\frac{1}{V} \Bigg( \frac{\partial V}{\partial \Pph} \Bigg)_{X_q} .
\end{align}
\end{subequations}

We also calculate the mean square of the fluctuation of the energy.
The quantity $ \<(\Delta \Uq{S})^2>$ is given by 
\begin{align}
  \<(\Delta \Uq{S})^2> = &(\TphS) \<(\Delta \Xq{S})^2> + (\PphS)^2 \< (\Delta V^{(S)})^2>
  - 2 \TphS \PphS \<(\Delta \Xq{S}) (\Delta V^{(S)})> + O(\Delta^3)  
\end{align}
The average $\<(\Delta \Xq{S}) (\Delta V^{(S)})>$ is given by
\begin{align}
  \<(\Delta \Xq{S}) (\Delta V^{(S)})>
  &\sim \left.\left( \frac{\partial \Xq{S}}{\partial V^{(S)}} \right)\right|_{\TphS} \<(\Delta V^{(S)})^2>
  = \left.\left( \frac{\partial \PphS}{\partial \TphS} \right)\right|_{V^{(S)}}\<(\Delta V^{(S)})^2>
  = \frac{\etaphS}{\kappaTS}\<(\Delta V^{(S)})^2>,
  \label{eqn:dXdV}
\end{align}
where we use the fact that the quantity $\< (\Delta \TphS) (\Delta V^{(S)})>$ is approximately zero.
The quantity $\eta_{\mathrm{ph}}$ in Eq.~\eqref{eqn:dXdV} is defined by
\begin{align}
  \eta_{\mathrm{ph}} = \frac{1}{V} \left( \frac{\partial V}{\partial \Tph} \right)_{\Pph} . 
\end{align}
With Eqs.~\eqref{fluctuations} and \eqref{eqn:dXdV}, we have
\begin{align}
  \<(\Delta \Uq{S})^2> \sim \Big( C_{qP}^{(S)} - 2\PphS V^{(S)} \etaphS \Big) (\TphS)^2 + \kappaTS (\PphS)^2 V^{(S)} \TphS.
  \label{eqn:dU2}
\end{align}

We calculate the quantity $\<(\Delta \Uq{S})^2>$ for ideal gas to check Eq.~\eqref{eqn:dU2}.
The energy and the equation of state for ideal gas are given by
\begin{subequations}
\begin{align}
  &U_q = \frac{3}{2} N \TphS, \label{ideal:U}\\
  &\PphS V^{(S)} = N \TphS. \label{ideal:eq-of-state}
\end{align}
\end{subequations}    
The quantity $\<(\Delta \Uq{S})^2>$ for ideal gas is obtained
from Eq.~\eqref{eqn:dU2} with Eqs.~\eqref{ideal:U} and \eqref{ideal:eq-of-state}:
\begin{align}
  \<(\Delta \Uq{S})^2> \sim \frac{3}{2} N (\TphS)^2 .
\end{align}
Therefore, we obtain the following ratio:
\begin{align}
  \frac{\sqrt{\<(\Delta \Uq{S})^2>}}{\Uq{S}} \sim \sqrt{\frac{2}{3N}} .
\end{align}
The ratio given above for ideal gas is well-known in the Boltzmann-Gibbs statistics.

We obtain the quantity $\< (\Delta \TphS)(\Delta \Xq{S})>$ for the calculation in the next subsection.
With Eq.~\eqref{eqn:CqV}, the quantity $\< (\Delta \TphS)(\Delta \Xq{S})>$ is given by
\begin{align}
  \< (\Delta \TphS)(\Delta \Xq{S})>
  = \left( \frac{C_{qV}^{(S)}}{\TphS} \right) \< (\Delta \TphS)^2>
  + \left(\frac{\partial \Xq{S}}{\partial V^{(S)}} \right) \< (\Delta \TphS) (\Delta V^{(S)}) >  + O(\Delta^3) .
\end{align}
We have
\begin{align}
  \< (\Delta \TphS)(\Delta \Xq{S})> \sim \left( \frac{C_{qV}^{(S)}}{\TphS} \right) \< (\Delta \TphS)^2>.
\label{eqn:dTdX}
\end{align}

The fluctuations of the physical quantities are obtained.
In the next subsection, we calculate the fluctuations of the Tsallis quantities, the Tsallis entropy and the Tsallis temperature.

\subsection{Fluctuations of the Tsallis quantities}
In this subsection, we estimate the fluctuations of the quantities appeared in the Tsallis statistics.
The fluctuation of the Tsallis entropy and the fluctuation of the Tsallis temperature are estimated in the following calculations. 
To proceed the calculations, we attempt to find the relations between the physical quantities and the Tsallis quantities.

The Tsallis temperature $\TS$ in the system is given by
\begin{align}
  \frac{1}{\TS} = \left( \frac{\partial S_q^{(S)}}{\partial U_q^{(S)}} \right)_{V^{(S)}} .
\end{align}
With Eq.~\eqref{def:Tph}, this equation leads to
\begin{align}
  \TphS = ( 1+(1-q)\Sq{S}) \TS = e^{(1-q) X_q^{(S)}} \TS . 
  \label{eqn:Tph-T}
\end{align}
The pressure $\PS$ in the system is defined by
\begin{align}
  \frac{\PS}{\TS} = \left(\frac{\partial S_q^{(S)}}{\partial V^{(S)}}\right)_{U_q^{(S)}} . 
  \label{def:Tsallis-Pressure}
\end{align}
From Eqs.~\eqref{def:Pph}, \eqref{eqn:Tph-T}, and \eqref{def:Tsallis-Pressure}, we have the relation $\PphS = \PS$.
Therefore, we focus on $\< (\Delta \Sq{S})^2>$ and $\< (\Delta \TS)^2>$.

It is possible to obtain the mean square of the relative fluctuation $\<(\Delta S_q/S_q)^2>$ for $q \neq 1$ from Eq.~\eqref{eq:fluc:X}
by using the relation between $\Xq{S}$ and $\Sq{S}$, Eq.~\eqref{eqn:relation:XS}, when $\Sq{S}$ is large enough.
The mean square of the fluctuation of the entropic variable, $\<(\Delta \Xq{S})^2>$, is represented with $\<(\Delta \Sq{S})^2 >$:
\begin{align}
  \< (\Delta \Xq{S})^2> = \Bigg(\frac{1}{1+(1-q)\Sq{S}}\Bigg)^2 \<(\Delta \Sq{S})^2> + O(\< (\Delta \Sq{S})^3>).
\end{align}
That is 
\begin{align}
\Bigg(\frac{1}{1+(1-q)\Sq{S}}\Bigg)^2 \<(\Delta \Sq{S})^2> \sim C_{qP}^{(S)} . 
\label{eqn:Cqp:Sq}
\end{align}
Equation~\eqref{eqn:Cqp:Sq} is also represented with $\Xq{S}$:
\begin{align}
  \<(\Delta \Sq{S})^2> \sim e^{2(1-q) \Xq{S}} C_{qP}^{(S)} .
\label{eqn:Cqp:Sq2}
\end{align}
For sufficiently large $\Sq{S}$ with $q \neq 1$, from Eq.~\eqref{eqn:Cqp:Sq}, we have
\begin{align}
\< (\Delta \Sq{S}/\Sq{S})^2> \sim (1-q)^2 C_{qP}^{(S)} . 
\end{align}
The quantity $(1-q)^2 C_{qP}^{(S)}$ gives the mean square of the relative fluctuation of the Tsallis entropy $\< (\Delta \Sq{S}/\Sq{S})^2>$
for sufficiently large $\Sq{S}$ with $q \neq 1$.

The mean square of the fluctuation of the Tsallis temperature $\< (\Delta \TS)^2>$ is calculated with Eq.~\eqref{eqn:Tph-T}. 
\begin{align}
  \< (\Delta \TS)^2>
  =  & e^{2(q-1) \Xq{S}} \Bigg\{ \< (\Delta \TphS)^2> + 2 (q-1) \TphS \< (\Delta \TphS)(\Delta \Xq{S})>  \nonumber \\
     & + (q-1)^2 (\TphS)^2 \< (\Delta \Xq{S})^2>  \Bigg\} + O(\Delta^3)  . 
\label{eqn:dT2}
\end{align}
Substituting Eq.~\eqref{eqn:dTdX} into Eq.~\eqref{eqn:dT2},
the quantity $\< (\Delta \TS)^2>$ is given by 
\begin{align}
  \< (\Delta \TS)^2>
  \sim  & e^{2(q-1) \Xq{S}} \Bigg\{ (1  + 2 (q-1) C_{qV}^{(S)}) \< (\Delta \TphS)^2> + (q-1)^2 (\TphS)^2 \< (\Delta \Xq{S})^2>  \Bigg\}.
\end{align}
The quantities, $\< (\Delta \TphS)^2>$ and $\< (\Delta \Xq{S})^2>$, are given by Eqs.~\eqref{eq:fluc:Tph} and \eqref{eq:fluc:X}.
The quantity $\< (\Delta \TS)^2>$ is represented as 
\begin{subequations}
\begin{align}
  \< (\Delta \TS)^2>
  & \sim  e^{2(q-1) \Xq{S}} (\TphS)^2 \Bigg\{ \left(\frac{1}{C_{qV}^{(S)}} \right) + 2 (q-1)  + (q-1)^2 C_{qP}^{(S)} \Bigg\} \\
  & = (\TS)^2 \Bigg\{ \left(\frac{1}{C_{qV}^{(S)}} \right) + 2 (q-1)  + (q-1)^2 C_{qP}^{(S)} \Bigg\}.
\end{align}
\end{subequations}
Therefore, the mean square of the relative fluctuation of the Tsallis temperature $\< ( \Delta \TS / \TS )^2 >$ is given by 
\begin{align}
  \< ( \Delta \TS / \TS )^2 >
  & \sim  \frac{1}{C_{qV}^{(S)}} + 2 (q-1)  + (q-1)^2 C_{qP}^{(S)} . 
\end{align}

The fluctuations of the Tsallis quantities in the Tsallis statistics approach the fluctuations in the Boltzmann-Gibbs statistics as $q$ approaches one.

\section{Conclusions}
In this paper, we considered the thermodynamic relations in the Tsallis statistics
by assuming the form of the first law with physical quantities.
We also calculated the fluctuations of thermodynamic quantities by using the entropic variable defined with the Tsallis entropy.

The first law is naturally described with the physical temperature $\Tph$ and the physical pressure $\Pph$
by introducing the entropic variable $X_q$ conjugate to the physical temperature.
The requirement that the first law has the form $dU_q = \Tph dX_q - \Pph dV$ suggests
the form of $X_q$ as a function of the Tsallis entropy $S_q$,
where $U_q$ is the energy and $V$ is the volume. 

We obtained the mean squares of the fluctuations of the physical quantities:
the physical temperature $\Tph$, the physical pressure $\Pph$, the volume $V$, and the entropic variable $X_q$.
The mean squares of the fluctuations of physical quantities in the Tsallis statistics are the same as those in the conventional statistics.
We also calculated the mean square of the fluctuation of the energy.
The mean square of the relative fluctuation of the energy for ideal gas has the well-known expression.

The mean square of the fluctuation of the Tsallis entropy $S_q$ and
the mean square of the fluctuation of the Tsallis temperature $T$ are given with physical quantities.
The mean square of the fluctuation of $S_q$ is represented with the heat capacity at constant (physical) pressure $C_{qP}$ and the entropic variable $X_q$.
The mean square of the fluctuation of $T$ is also represented with the physical temperature, heat capacities, and the entropic variable $X_q$.
The mean square of the relative fluctuation of the Tsallis entropy $\< ( \Delta S_q / S_q )^2 >$ with $q \neq 1$ is given by $(q-1)^2 C_{qP}$ for large $S_q$.
The mean square of the relative fluctuation of the Tsallis temperature $\< ( \Delta \TS / \TS )^2 >$
is represented with the deviation $(q-1)$ and heat capacities.
The fluctuations of the Tsallis quantities in the Tsallis statistics approach the fluctuations in the Boltzmann-Gibbs statistics
as the entropic parameter $q$ approaches one.

It is possible to compare the results in this study with the results in other studies.
The thermodynamic quantities in the R\'enyi statistics, in which the standard average is employed, were investigated,
and it was shown that the thermodynamic quantities in the R\'enyi statistics are the same as those in the Boltzmann-Gibbs statistics \cite{Parvan:2010}.
It was also shown that the physical quantities are $q$-independent within the framework of Tsallis statistics \cite{Toral:2003}.
This property is shown in the present study:
the mean squares of the fluctuations of the physical quantities, $\Tph$, $V$, $X_q$, and $\Pph$, are the same as those in the Boltzmann-Gibbs statistics.
The fluctuations of the physical quantities do not contain the entropic parameter $q$ explicitly:
the deviation $q-1$ does not appear.
In contrast, the fluctuation of the Tsallis entropy $S_q$ and
the fluctuation of the Tsallis temperature $T$ contain the entropic parameter $q$ explicitly:
the deviation $q-1$ appears in the fluctuations.

The fluctuations of the physical quantities do not contain the entropic parameter $q$ explicitly,
and the fluctuations are the same as those in the Boltzmann-Gibbs statistics.
This fact comes from the relation between the entropic variable $X_q$ and the number of states $W$: $X_q = \ln W$.
The fluctuation of the Tsallis entropy $S_q$ and
the fluctuation of the Tsallis temperature $T$ contain the entropic parameter $q$ explicitly.
This fact comes from the relation between the entropy $S_q$ and the number of states $W$. 
The entropic parameter appears explicitly in the relation: $S_q = \ln_q W$.

The discussion above for the Tsallis entropy is extended to the discussions for other entropies.
We have the equation, $\ln W = \ln( f_{\mathrm{inv}}(S))$, 
when an entropy $S$ and the number of states $W$ have the relation $W = f_{\mathrm{inv}} (S)$,
where $f_{\mathrm{inv}}$ is the inverse function of a function $f$.
The fluctuation of the physical variable is the same as the fluctuation of the corresponding variable in the Boltzmann-Gibbs statistics, 
when the physical variable is defined by using $\ln(f_{\mathrm{inv}}(S))$.
However, as shown in this paper, the fluctuation of the variable defined directly from the entropy $S$ (for example, Tsallis entropy)
is not the same as the fluctuation of the corresponding variable in the Boltzmann-Gibbs statistics:
the fluctuation of the Tsallis temperature is not the same as the fluctuation of the temperature in the Boltzmann-Gibbs statistics.

The physical quantities such as physical temperature characterize the equilibrium.
Therefore it might be better to use the name, equilibrium temperature, instead of physical temperature.
It was pointed out that two temperatures are not the same for two finite systems \cite{Prosper1993}.
Therefore, we may pay attention to the observed temperature.
Whether the physical temperature is the observed temperature depends on the property of the thermometer.

The calculations of the fluctuations are based on the expression, $W = \exp(X_q)$. 
That is, the calculations depend on the property of the exponential function: $\exp(x+ \Delta x) = \exp(x) \exp(\Delta x)$.
The calculations also depend on the division between the system and the environment represented as $W^{(S+E)} = W^{(S)} W^{(E)}$, 
where $W^{(S+E)}$, $W^{(S)}$, and $W^{(E)}$ are the number of states for the whole system, that for the system, and that for the environment.
Therefore, the results in the present study will be modified in the case of $W^{(S+E)} \neq W^{(S)} W^{(E)}$. 

In this paper, the thermodynamic relations were studied in the Tsallis statistics.
The mean squares of the fluctuations of the physical quantities and
the mean squares of the fluctuations of the Tsallis quantities were obtained. 
The results given in this paper will be helpful to study the phenomena described with the statistics related to power-like distributions.

\medskip\noindent\textbf{Funding} This research received no specific grant from any funding agency in the public, commercial, or not-for-profit sectors.

\medskip\noindent\textbf{Data availability statement}
This manuscript has no associated data or the data will not be deposited.
[Authors' comment: This study is theoretical, and no data is generated.]

\medskip\noindent\textbf{Competing Interests} The author declares no competing interest.


\end{document}